\documentstyle[aps]{revtex}
\begin{document}
\newcommand{\beq}{\begin{equation}}
\newcommand{\eeq}{\end{equation}}
\newcommand{\beqn}{\begin{eqnarray}}
\newcommand{\eeqn}{\end{eqnarray}}
\newcommand{\bmath}{\begin{mathletters}}
\newcommand{\emath}{\end{mathletters}}
\twocolumn[\hsize\textwidth\columnwidth\hsize\csname @twocolumnfalse\endcsname
\title{Dynamic Hubbard Model}
\author{J. E. Hirsch }
\address{Department of Physics, University of California, San Diego\\
La Jolla, CA 92093-0319}
 
\date{July 24, 2001} 
\maketitle 
\begin{abstract} 
The Hubbard on-site repulsion $U$ between opposite spin electrons on the same
atomic orbital is widely regarded to be the most important source of electronic
correlation in solids. Here we extend the Hubbard model to account for the fact that
the experimentally measured atomic $U$ is different from the one obtained
by calculation of the atomic Coulomb integral. The resulting model describes 
quasiparticles that become increasingly dressed as the number of electrons in the
band increases. Superconductivity can result in this model through quasiparticle undressing.
Various signatures of this physics in spectroscopies in the normal and
superconducting states are discussed. A novel effect 
in the normal state is predicted to be electroluminescence
at the sample-positive counterelectrode boundary.
\end{abstract}
\vskip2pc]

The Coulomb repulsion integral for two electrons in a hydrogenic 1s orbital
for a nucleus of charge $Z$ is
\bmath
\beqn
U(Z)&=&\int d^3r d^3 r' |\varphi _{1s}(r)|^2 \frac{e^2}{|r-r'|}|\varphi _{1s}(r')|^2
\nonumber \\
&=&\frac{5}{4}Z\times 13.606 eV
\eeqn
\beq
\varphi _{1s}(r)=(\frac{Z^3}{\pi a_0^3})^{1/2} e^{-Zr/a_0}
\eeq
\emath
($a_0=$Bohr radius). However, the effective atomic on-site repulsion 
$U_{eff}(Z)$ is obtained
by considering the difference in energy between ions with different number
of electrons. For Hydrogen ($Z=1$) and Helium ($Z=2$) the experimental values 
are\cite{hand}
\bmath
\beq
U_{eff}(1)=I-A=12.86 eV=U(1)-4.15 eV
\eeq
\beq
U_{eff}(2)=I_{II}-I_I=29.92 eV=U(2)-4.10 eV
\eeq
\emath
Here, $I=13.6eV$, $A=0.747eV$, $I_I=24.48 eV$, $I_{II}=54.40 eV$ are the ionization
energy and electron affinity of $H$, and the first and second ionization
energies of $He$. Remarkably, the difference between $U_{eff}$ and $U$ is
nearly the same for $Z=1$ and $Z=2$. Indeed, the same is true for higher
$Z$, for $Z$ between 3 and 8 the difference
between $U$ and $U_{eff}$ for the 1s atomic orbital, obtained from
the appropriate ionization energies\cite{hand} is respectively , in eV, 
$4.22$, $4.22$, $4.21$, $4.19$, $4.15$, $4.05$. 
We conclude that for $1s$ orbitals one has approximately
\beq
U_{eff}(Z)=U(Z)-4.1 eV .
\eeq
for a large range of $Z$. 
A similar reduction from bare $U$ to effective $U$ will be found for other
atomic orbitals. This reduction occurs because when the second electron
is added to the singly occupied orbital the state of the two-electron system
is not the doubly occupied single electron orbital. Rather, the orbital will
'expand' to reduce the Coulomb repulsion between electrons, and furthermore
the two electrons will develop angular correlations. This effect is of course well
known in atomic physics, and in its simplest form is approximately described
by Slater's rules for the shielding constants\cite{sla}: when another electron is added
to an atom, the effective $Z$ for the electrons in the same shell is reduced.
This effect is however ignored in the ordinary Hubbard model\cite{hub}.

A simple way to describe this physics is by introducing coupling to a 
fictitious local boson displacement coordinate $q_i$ for atom $i$ that modulates
the Hubbard $U$:
\beq
U(q_i)=U+\alpha q_i
\eeq
that will relax when double occupancy occurs. As the simplest model we 
describe the boson dynamics by a harmonic oscillator of frequency $\omega_0=(K/M)^{1/2}$:
\beq
H_i=\frac{p_i^2}{2M} + \frac{1}{2} K q_i^2 + (U+\alpha q_i)n_{i\uparrow}n_{i\downarrow}
\eeq
and the effective on-site repulsion is found by completing the squares as
$U_{eff}=U-\alpha^2/2K$. The equilibrium position of the boson is $q_i=0$ for
the orbital empty or singly occupied, and $q_i=-\alpha/K$ for the doubly
occupied orbital. 

The reader may argue that the ordinary Holstein model\cite{holst}
\beq
H_i=\frac{p_i^2}{2M} + \frac{1}{2} K q_i^2 +\alpha q_i(n_{i\uparrow}+n_{i\downarrow})
+Un_{i\uparrow}n_{i\downarrow}
\eeq
will also describe a reduction of the bare $U$ to a $U_{eff}=U-\alpha^2/K$.
However, in contrast to Eq. (5), Eq. (6) also describes dressing of electrons in
$singly$-occupied orbitals by the boson degree of freedom. That is not the physics we
are trying to describe here: without electron-phonon interactions or coupling
to atomic electrons in other orbitals, the electron in the singly occupied
orbital should be undressed. 

In terms of boson creation and annihilation operators, the Hamiltonian Eq. (5) is
\bmath
\beq
H_i=\omega_0 a_i^\dagger a_i+[U+g\omega_0(a_i^\dagger+a_i)]n_{i\uparrow}n_{i\downarrow}
\eeq
\beq
U_{eff}=U-w_0g^2
\eeq
\emath
with $g=\alpha/(2K \omega_0)^{1/2}$. The boson degree of freedom describes the
electronic excitation of an electron when a second  electron is added to the
orbital. Hence the frequency $\omega_0$ is related to the
excitation energies of the atom, and we expect
\bmath
\beq
\omega_0=cZ^2
\eeq
with c a constant of order eV, since the excitation energies in an atom scale with
the square of the nuclear charge. From Eqs. (3) and (7b) we conclude
\beq
g^2=\frac{c'}{Z^2}
\eeq
\emath
with $c'=4.1 eV/c$. For a lattice system where an electron hops from site $i$ to
site $j$ with hopping amplitude $t_{ij}$ the Hamiltonian is then
\beqn
H&=&-\sum_{ij,\sigma}t_{ij} (c_{i\sigma}^\dagger c_{j\sigma} +h.c.)
+
\sum_i[U+g\omega_0(a_i^\dagger +a_i)]n_{i\uparrow}n_{i\downarrow}
\nonumber \\ & &
+\sum_i\omega_0 a_i^\dagger a_i
\eeqn

The physics described by Eq. (3), that led to the Hamiltonian Eq. (9),
is a ubiquitous phenomenon, originating in the fact that the
spacing between atomic energy levels is always smaller than the strength of the
Coulomb repulsion between electrons in a given orbital. Its quantitative importance
is determined by the magnitude of the ionic charge $Z$, as discussed
below. Other models, with auxiliary spin degrees of freedom\cite{spin} or with only
electronic degrees of freedom\cite{elec}, can be constructed containing the same
physics. The model Eq. (9) is a particular case of the generalized Holstein
models discussed in ref. \cite{undr}.

By construction, the model Eq. (9)  is not electron-hole symmetric, even if the 
band structure defined by $t_{ij}$ is. In particular, a single electron in
the empty band does not interact with the boson field at all. 
For a few electrons in an empty band the effect of the
bosons is negligible if the bare Coulomb repulsion $U$ is appreciable,
as the probability of double occupancy will be small. In contrast, a
single hole in the full band interacts most strongly with the boson field.
Treating the four-fermion term in mean field, the electron-boson
part of the Hamiltonian Eq. (9) is
\bmath
\beq
H_{el-b}=g(n)\omega _0 (a_i^\dagger + a_i) (n_{i\uparrow}+n_{i\downarrow})
\eeq
\beq
g(n)=\frac{n}{2}g
\eeq
\emath
that is, an ordinary electron-boson coupling with a coupling constant that increases
monotonically with band filling. Hence, as the usual electron-phonon 
interaction, it will give rise to an effective mass enhancement and a 
quasiparticle weight reduction, which increase as the band filling and hence the
effective coupling $g(n)$ increases. There are however other dynamical
effects of Eq. (9) that are lost in the mean field treatment Eq. (10).

To study Eq. (9) we perform a generalized Lang-Firsov transformation on the
fermion and boson operators\cite{lang1,undr}
\bmath
\beq
c_{i\sigma}=e^{g(a_i^\dagger-a_i)\tilde{n}_{i,-\sigma}}
\tilde{c}_{i\sigma}\equiv X_{i\sigma}\tilde{c}_{i\sigma}
\eeq
\beq
a_i=\tilde{a}_i-g\tilde{n}_{i\uparrow}\tilde{n}_{i\downarrow}
\eeq
\emath
and the Hamiltonian Eq. (9) becomes
\beqn
H=&-&\sum_{ij,\sigma}t_{ij} (X_{i\sigma}^\dagger X_{j\sigma}
\tilde{c}_{i\sigma}^\dagger \tilde{c}_{j\sigma} +h.c.)
+
\nonumber \\
& &\sum_i U_{eff} \tilde{n}_{i\uparrow}\tilde{n}_{i\downarrow}
+\sum_i\omega_0 \tilde{a}_i^\dagger \tilde{a}_i
\eeqn
with $U_{eff}$ given by Eq. (7b).  The ground
state expectation value of the $X_{i\sigma}$ operator
is
\bmath
\beq
<X_{i\sigma}>_0=e^{-(g^2/2) \tilde{n}_{i,-\sigma}}= 1+(S-1)\tilde{n}_{i,-\sigma}
\eeq
\beq
S=e^{-g^2/2}
\eeq
\emath
The part of the fermion operator Eq. (11a) associated with ground state to
ground state transitions of the boson field is the coherent part of the
operator, the quasiparticle. We have then
\beq
c_{i\sigma}=|0><0|[1+(S-1) \tilde{n}_{i,-\sigma}]\tilde{c}_{i\sigma}
+c_{i\sigma}^{incoh}
\eeq
The incoherent part of the operator
\beqn
c_{i\sigma}^{incoh}&=&[\tilde{n}_{i,-\sigma}\sum_{(l,l')\neq(0,0)}|l><l|
e^{g(a_i^\dagger-a_i)}|l'><l'| \nonumber \\
& &+\sum_{l\neq 0}|l><l|]\tilde{c}_{i\sigma}
\eeqn
describes processes where the boson field makes transitions to and from
excited states $|l>, l\neq 0$, which only take place if $\tilde{n}_{i,-\sigma}=1$,
that is if the orbital is occupied by another electron of opposite spin.

	The quasiparticle weight in this model is, from Eq. (14)
\beq
z(n)=[1+\frac{n}{2}(S-1)]^2
\eeq
and decreases monotonically with electronic band filling $n$, $0\leq n\leq 2$, so that
quasiparticles become increasingly dressed as the band filling increases.
The factor $S$ is the overlap matrix element of the oscillator ground states with
and without site double occupancy\cite{undr}, and $S^2$ gives the quasiparticle weight
for a hole in the filled band ($n=2$ in Eq. (16)). According to Eq. (8b), as the ionic 
charge $Z$ decreases $S$ decreases rapidly, implying that hole quasiparticles become
increasingly incoherent. 

We can  estimate $S$  from first principles
for a hydrogen-like ion. In  the Hartree approximation, $S$ will be given by the
overlap matrix element of the electron wave function in the presence 
 and in the absence of another electron in the orbital:
\beq
S=|<\varphi_{1s}|\bar{\varphi}_{1s}>|=
\frac{(1-\frac{5}{16Z})^{3/2}}{(1-\frac{5}{32Z})^3}
\eeq
with $\bar{\varphi}_{1s}$ the 1s orbital Eq. (1b) with $Z$ replaced by 
$\bar{Z}=Z-5/16$, as appropriate for the Hartree wavefunction. If we use the
more accurate Eckart wave function\cite{eck1,eck2} $\psi_{Eck}$ for the
two-electron ion, which incorporates radial correlations, we can estimate $S$
from the square root of the overlap matrix element of the Eckhart wave function
and the wavefunction of the two electrons in the $1s$ orbital, as
\bmath
\beq
S=\sqrt{|<\varphi_{1s}\varphi_{1s}|\psi_{Eck}>|^2}=\sqrt{\frac{2f(Z,Z_1)f(Z,Z_2)}
{(2(1+f(Z_1,Z_2))^{1/2}}}
\eeq
\beq
f(Z,Z')=\frac{(Z Z')^{3/2}}{(\frac{Z+Z'}{2})^3} .
\eeq
\emath
Here, $Z_1$ and $Z_2$ are the two orbital exponents for the Eckart wave function
obtained by minimization of the energy, which are found to be
approximately $Z_1=1.14Z-0.105$, $Z_2=0.905 Z-0.622$. The
Eckart wave function becomes unstable for $Z<0.93$.

Figure 1 shows the dependence of $S$ on the ionic charge $Z$. We expect the
Eckart wavefunction to underestimate and the Hartree wavefunction to 
overestimate the value of $S$ for a given $Z$. Figure 1 also shows $S$ for the
dynamic Hubbard model, Eqs. (13b) and (8b), for a value of $c'$ that matches
the Eckart wavefunction results for large $Z$. While the three curves are
different, the qualitative behavior is the same, showing a decrease of $S$ as
the ionic charge $Z$ decreases, which describes the increased incoherence of 
single hole carriers in the system as the ionic charge decreases.

Upon replacement of the operator form Eq. (14) in the Hamiltonian Eq. (12),
and ignoring the incoherent part of the operators, we obtain 
 the effective Hamiltonian describing propagation of 
quasiparticles:
\bmath
\beq
H_{qp}=-\sum_{ij\sigma}\tilde{t}_{ij}^\sigma (\tilde{c}_{i\sigma}^\dagger
\tilde{c}_{j\sigma}+h.c.) +\sum_i U_{eff}\tilde{n}_{i\uparrow}\tilde{n}_{i\downarrow}
\eeq
where the hopping amplitude now will depend on the occupation of the two sites
involved in the hopping process:
\beq
\tilde{t}_{ij}^\sigma=t_{ij}[1+(S-1)(\tilde{n}_{i,-\sigma}+\tilde{n}_{j,-\sigma})
+(S-1)^2 \tilde{n}_{i,-\sigma}\tilde{n}_{j,-\sigma}]
\eeq
\emath
The effective quasiparticle Hamiltonian Eq. (19) will accurately describe the
low energy physics of the full Hamiltonian Eq. (12) in the strong coupling
regime, where the Lang-Firsov approximation to the Holstein model becomes
accurate. Even in that regime however the high energy degrees of freedom 
described by the full Hamiltonian Eq. (12) still play an important role, in
ensuring that various sum rules that are violated by the low energy Hamiltonian
Eq. (19a) are satisfied.

The hopping amplitude for an electron of spin $\sigma$ when there are
electrons of opposite spin at both sites $i$ and $j$, i.e.
$\tilde{n}_{i,-\sigma}=\tilde{n}_{j,-\sigma}=1$ is, from Eq. (19b),
$t_2=S^2t_{ij}$, and when there is one other electron at the two sites, i.e.
$\tilde{n}_{i,-\sigma}+\tilde{n}_{j,-\sigma}=1$,
$t_1=St_{ij}$. 
If $S<<1$ then $t_2<<t_1$. In the limit where $t_2$ can be neglected
with respect to $t_1$ it becomes obvious that the Hamiltonian Eq. (19)
leads to pairing of two holes in a full band: a state where the two
holes are separate describes essentially localized holes with zero 
energy, since $t_2\sim0$; a lower energy state results from a linear
combination of states where the two holes are on the same or
nearest neighbor sites, with energy of order
$\epsilon_{pair}\sim -2p\times t_1^2/U_{eff}$
with $p$ the number of nearest neighbors to a site. More generally, an
exact criterion can be found for the parameters that will yield a bound
state for two holes in a filled band described by the Hamiltonian
Eq. (19)\cite{exact}:
\beq
\frac{U_{eff}}{D}\leq 1-S^2
\eeq
where $D$ is the bare bandwidth of the band defined by $t_{ij}$.
The same criterion is found for existence of superconductivity in the
dilute carrier concentration regime in BCS theory\cite{hole}, a regime where BCS
theory is expected to be accurate, and quantitatively close results are 
obtained from exact diagonalization of small systems\cite{lin}. As the
atomic charge $Z$ decreases, the coupling $g$ increases (Eq. (8b)), hence
$S$ decreases (Eq. (13b)), and furthermore $U_{eff}$ decreases
(Eqs. (3) and (1a)). Both of these effects are in the right direction to 
satisfy the condition Eq. (20). Furthermore, if the interatomic
distance is decreased, $t_{ij}$ and the bare bandwidth $D$ will increase,
again in the direction of satisfying the inequality Eq. (20). 
We conclude that superconductivity induced by this physics will occur in 
systems where the ionic charge $Z$ is small, the interatomic distances are
small, and the Fermi level is close to the top of a band. The parameter regime where
the Hamiltonian Eq. (9) yields superconductivity has not yet been
established, however numerical studies of the similar spin-1/2 Hamiltonian\cite{spin}
have shown that pairing survives well beyond the parameter regime
where the mapping to the effective Hamiltonian Eq. (19) is accurate.

The dressing of hole quasiparticles decreases in this model as the
hole concentration increases, and correspondingly the hole quasiparticle
weight increases, as seen from Eq. (16) or its equivalent in
hole representation
\beq
z_h(n_h)=S^2[1+\frac{n_h}{2}\Upsilon]
\eeq
with $n_h=2-n$, $\Upsilon=1/S-1$. In the normal state, this should be 
seen in the one-particle Green's function as a transfer of spectral
weight from the incoherent part, describing high energy excitations of the
boson field on the scale of $\omega_0$, to the quasiparticle peak as the
hole doping increases. 
Correspondingly, a transfer of spectral weight in optical absorption
should be seen, from high frequency absorption on the scale of
$\omega_0$ to intra-band Drude absorption with plasma frequency
determined by $t_2(n_h)=t_2+n_h(t_1-t_2)$. From the operator relationship for
hole quasiparticles derived from Eq. (14)
\beq
c_{i\sigma}^{h\dagger}=S[1+\Upsilon \tilde{n}_{i,-\sigma}^h]\tilde{c}_{i\sigma}^{h\dagger}
\eeq
it is  seen that the expectation value $<c_{i\sigma}^h c_{j\sigma}^{h\dagger}>$
acquires contributions from anomalous expectation values 
$<\tilde{c}_{i\sigma}^h \tilde{c}_{j\sigma}^h>$.
This will cause transfer of spectral weight from high to low frequencies
in the one- and two-particle Green's functions for fixed hole doping when the system goes
superconducting, which should also be observable in photoemission and
optical absorption experiments\cite{undr}.

Non-perturbative theoretical techniques such as dynamical mean field 
theory\cite{dynmf} and DMRG\cite{dmrg} should be able to establish
the parameter regime in the Hamiltonian Eq. (9) where the physics of
superconductivity through undressing described above takes place. 
First-principles quantum chemical and density-functional calculations
should be able to relate the parameters in Eq. (9) to real materials.
The physics discussed here predicts that superconductivity is favored 
in systems where conduction is through  holes in nearly filled bands, through
conducting structures that are negatively charged (small ionic charge $Z$),
and that such systems will show undressing of hole carriers both when the system 
is doped with 
holes and when it goes superconducting. These effects will be most
apparent when $Z$ is small and $\Upsilon$ is large, which also leads to high $T_c$.
The facts that most superconductors show hole carrier transport in the
normal state\cite{chap}, that both $MgB_2$ and the high $T_c$ cuprates have
holes conducting in highly negatively charged substructures 
($B^-$ and $(CuO_2)^=$  planes), and that
the high $T_c$ cuprates show evidence of 'undressing' upon hole
doping and upon going superconducting\cite{photo,opt,mass,undr}, suggest that the
physics of the dynamical Hubbard model may have something to do
with the physics of superconductivity in real materials.

In the regime most favorable for superconductivity, i.e. hole conduction in a 
system with large $\Upsilon$, the model Eq. (9) predicts that incoherent
excitations on electronic energy scales ($\omega_0$) will be induced
when holes hop. Hence one would expect non-thermal high frequency radiation
to be generated when a dc current circulates in the normal state. 
This electroluminescence\cite{elum}
should be most pronounced and easily observable at the positive counterelectrode-sample
boundary, where holes are injected into the sample. The 
intensity of the radiation should correlate with the magnitude of the dressing
and hence of $T_c$, and the frequency distribution
 will give information on the scale of electronic excitation
energies involved in the undressing process that leads to superconductivity.

\begin{figure}
\caption {Parameter $S$ versus ionic charge $Z$ in the Hartree
and Eckart approximations, and in the electron-boson model with
$c'=0.2$ (Eqs. (8b), (13b)). The deviation of $S$ from $1$
indicates the degree of quasiparticle dressing for an
almost full band.
}
\label{Fig. 1}
\end{figure}


\begin{references}
\bibitem{hand} Handbook of Chemistry and Physics, 59-th Edition, 
CRC Press, West Palm Beach, 1978.
\bibitem{sla} J.C. Slater, {\it Quantum Theory of Atomic Structure},
McGraw Hill, New York, 1960, Chpt. 15.
\bibitem{hub} J.E. Hirsch, Physica B {\bf 199\&200}, 366 (1994).
\bibitem{holst} T. Holstein, Ann.Phys. N.Y. {\bf 8}, 325 (1959).
\bibitem{spin} J.E. Hirsch and S. Tang, Phys. Rev. B {\bf 40}, 2179 (1989);
J.E. Hirsch, cond-mat/0109385 (2001) .
\bibitem{elec} J.E. Hirsch, Phys. Rev. B {\bf 43}, 11400 (1991).
\bibitem{undr}  J.E. Hirsch, Physica C {\bf 201}, 347 (1992);  
{\bf 364-365}, 37 (2001); Phys. Rev. B {\bf 62}, 14487 (2000); {\bf 62 }, 14498 (2000).
\bibitem{lang1} I.G. Lang and Y.A. Firsov, Zh. Eksp. Teor. Fiz. {\bf 43},
923 (1962) [Sov. Phys. JETP {\bf 16}, 1301 (1963)].
\bibitem{eck1} C. Eckart, Phys. Rev. {\bf 36}, 878 (1930).
\bibitem{eck2} J.E. Hirsch, Phys. Rev. B {\bf 48}, 3340 (1993).
\bibitem{exact} J.E. Hirsch, Phys. Rev. B {\bf 47}, 5351 (1993).
\bibitem{hole} J.E. Hirsch and F. Marsiglio, Phys. Rev. B {\bf 39}, 11515
(1989); ;  Physica C {\bf 162-164}, 591 (1989); F. Marsiglio and J.E. Hirsch,
Phys. Rev. B {\bf 41}, 6435 (1990).
\bibitem{lin} H.Q. Lin and J.E. Hirsch, Phys. Rev. B {\bf 52}, 16155 (1995).
\bibitem{dynmf} A. Georges et al, Rev. Mod. Phys. {\bf 68}, 13 (1996).
\bibitem{dmrg} C. Zhang, E. Jeckelmann and S.R. White, 
Phys. Rev. B {\bf 60}, 14092 (1999).
\bibitem{chap} I.M. Chapnik, Sov.Phys. Dokl. {\bf 6}, 988 (1962).
\bibitem{photo} H. Ding et al, cond-mat/0006143 (2000).
\bibitem{opt} S. Uchida et al, Phys. Rev. B {\bf 43}, 7942 (1991);
I. Fugol et al, Sol.St.Comm. {\bf 86}, 385 (1993); D. van der Marel et al, 
private communication.
\bibitem{mass} Y. Ando et al, Phys. Rev.Lett. {\bf 87}, 017001 (2001);
P.D. Johnson et al, Phys. Rev.Lett. {\bf 87}, 177007 (2001) .
\bibitem{elum} H.F. Ivey, ``Electroluminescence and Related Effects'',
Academic Press, New York, 1963.


\end{references}
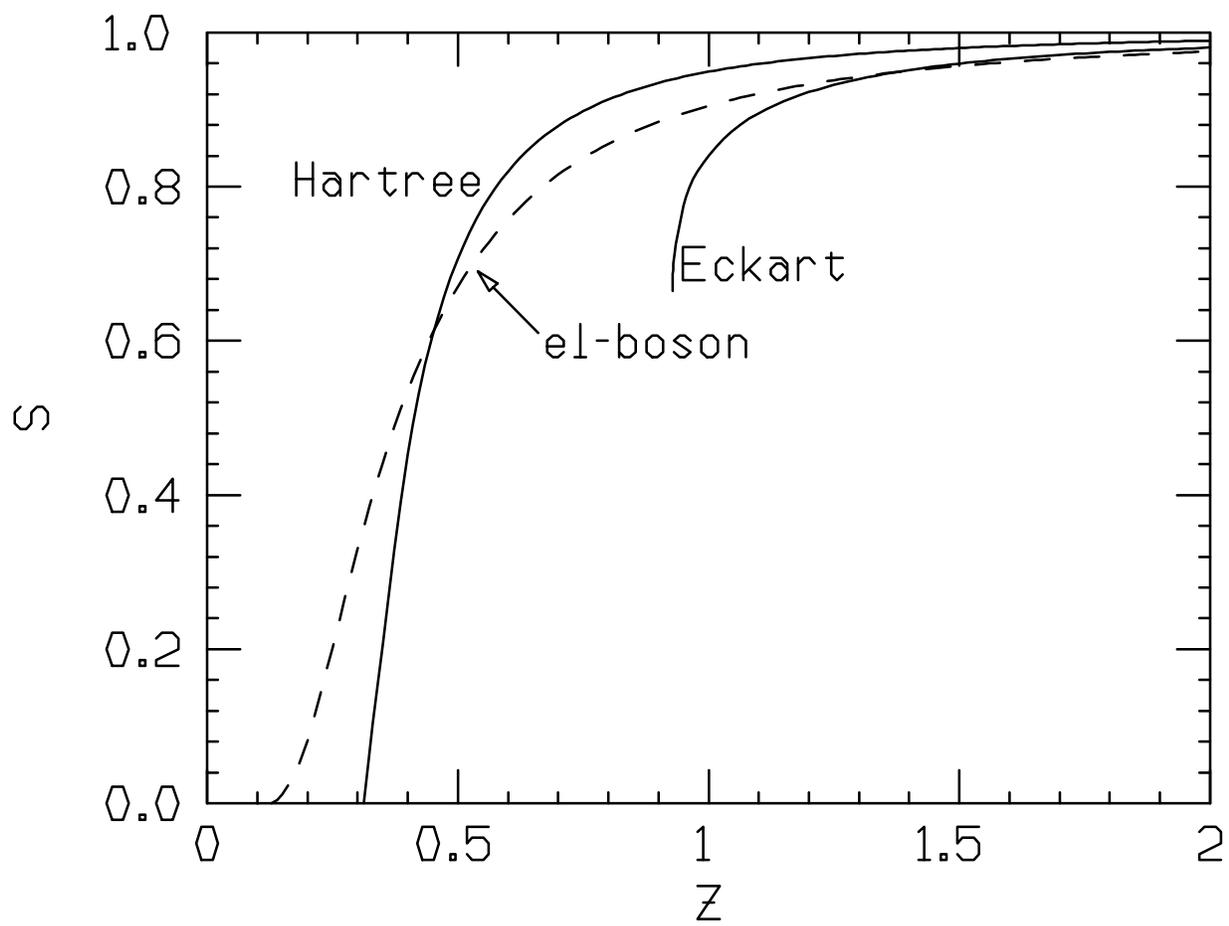
\end{document}